\documentclass[aps,12pt]{revtex4}
\usepackage{graphicx}
\usepackage{wasysym}
\newcommand{\bm}{\boldmath}
\newcommand{\D}{\displaystyle}
\newcommand{\vek}[1]{\mbox{\bm ${#1}$}}
\begin{document}
\thispagestyle{empty}
\renewcommand{\theequation}{\arabic{section}.\arabic{equation}}
\begin{center}
{\bf Corrected Draft 8 September 2006} 
\end{center}
\vspace{3cm}
\begin{center}{\large\bf The on-shell self-energy of the uniform electron gas in its weak-correlation limit}\\
\vspace{2cm}
{\sc P. Ziesche} \\
\vspace{1cm}
{Max-Planck-Institut f\"ur Physik komplexer Systeme \\
N\"othnitzer Str. 38, D-01187 Dresden, Germany}\\
\vspace{1cm}
\date{\today}
\vspace{1cm}
\end{center}

\vspace{2cm}

\begin{abstract}
\noindent
The ring-diagram partial summation (or RPA) for the ground-state energy of the uniform electron gas (with the 
density parameter $r_s$) in its weak-correlation limit $r_s\to 0 $ is revisited. It is studied, which 
treatment of the self-energy $\Sigma(k,\omega)$ is in agreement with the Hugenholtz-van Hove 
(Luttinger-Ward) theorem $\mu-\mu_0= \Sigma(k_{\rm F},\mu)$ and which is not. The correlation part of the lhs has the 
RPA asymptotics
$a\ln r_s +a'+O(r_s)$ [in atomic units]. The use of renormalized RPA diagrams for the rhs yields the similar 
expression $a\ln r_s+a''+O(r_s)$ with the sum rule $a'= a''$ resulting from three sum rules for the components of
$a'$ and $a''$. This includes in the second order of exchange the 
sum rule $\mu_{2{\rm x}}=\Sigma_{2{\rm x}}$ [P. Ziesche, Ann. Phys. (Leipzig), 2006].  
\end{abstract}
\maketitle
\newpage
\section{Introduction}
\noindent
Although not present in the Periodic Table the uniform or homogeneous electron gas (HEG) is still an important 
model system for electronic structure theory, cf. e.g. \cite{Tos}. In its spin-unpolarized version, the 
HEG ground state is characterized by only one parameter $r_s$, such that a sphere with the radius 
$r_s$ contains {\it on average} one electron \cite{Zie1}. It determines the Fermi wave number as 
$k_{\rm F}=1/(\alpha r_s)$ in a.u. with $\alpha =[4/(9\pi)]^{1/3}\approx 0.521062$ and it measures simultaneously 
both the interaction strength and the density such that high density corresponds to weak interaction and 
hence weak correlation \cite{foo}. For recent papers on this limit cf. \cite{Zie2,Zie3,Cio,Mui}. Usually the 
total ground-state energy per particle is written as (here and in the following are wave numbers measured in units of 
$k_{\rm F}$ and energies in $k_{\rm F}^2$)
 \begin{equation}\label{a1}
e=e_0+e_{\rm x}+e_{\rm c}, \quad e_0=\frac{3}{5}\cdot \frac{1}{2}, \quad 
e_{\rm x}=-\frac{3}{4}\cdot\frac{\alpha r_s}{\pi}, \quad e_{\rm c}=(\alpha r_s)^2[a\ln r_s +b+b_{2{\rm x}}+O(r_s)],
\end{equation}  
where $e_0$ is the energy of the ideal Fermi gas, $e_{\rm x}$ is the exchange energy in lowest (1st) order,
and $e_{\rm c}$ is referred to as correlation energy given here in its weak-correlation limit with 
$a=(1-\ln 2)/\pi^2\approx 0.031091$ after Macke \cite{Ma} and $b\approx -0.0711$ after Gell-Mann and Brueckner 
\cite{GB}. $e_{\rm c}$ contains also the 2nd-order of exchange with $b_{2{\rm x}}\approx +0.02418$ after
Onsager, Mittag, and Stephen \cite{Ons}. Notice that $\tilde e=k_{\rm F}^2e=e/(\alpha r_s)^2$ gives the energy in a.u., 
e.g. the energy in zeroth order and the lowest-order exchange energy are ${\tilde e}_0=3/(10\ \alpha^2 r_s^2)$ and 
${\tilde e}_{\rm x}=-3/(4 \pi \alpha r_s)$, respectively. \\ 

\begin{figure}[h!]
\begin{center}
\rotatebox{0}{\resizebox{75mm}{!}{\includegraphics{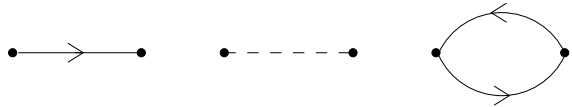}}}
\caption{\label{fey1}Feynman diagrams for the one-body Green's function of the ideal Fermi gas $G_0(k,\omega)$, 
the bare Coulomb repulsion $v_0(q)$, and the RPA polarization propagator $Q(q,\eta)$ as defined in 
Eqs. (\ref{A2})-(\ref{A4}).} 
\end{center}
\end{figure}

\begin{figure}[h!]
\begin{center}
\rotatebox{0}{\resizebox{100mm}{!}{\includegraphics{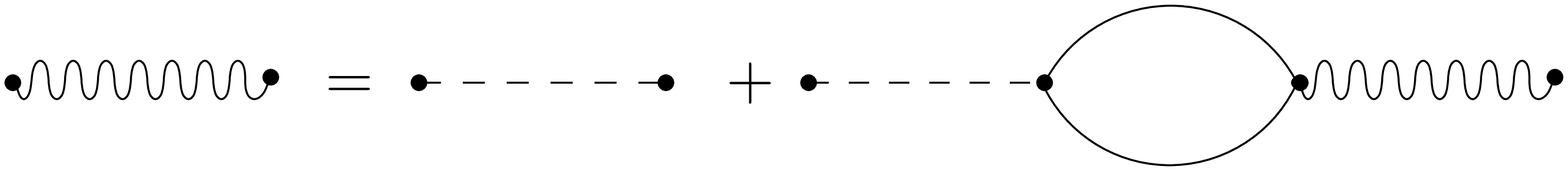}}}
\caption{\label{fey2}Feynman diagrams for $v_{\rm r}=v_0+v_0Qv_{\rm r}$, the screened Coulomb repulsion in RPA.}
\end{center}
\end{figure}

\noindent
Revisiting how Macke \cite{Ma}, Gell-Mann/Brueckner \cite{GB}, and Onsager/Mittag/Stephen \cite{Ons,Gla} derived 
$e_{\rm c}$ in its weak-correlation limit, it is shown here, that and how an analogous  
procedure - also called RPA (= random phase approximation) - applies to the self-energy $\Sigma(k,\omega)$. This latter quantity determines 
(i) the one-body Green's function $G(k,\omega)$, from which follow the quasi-particle dispersion and damping and 
the momentum distribution $n(k)$ \cite{GGZ}. It furthermore appears (ii) in the Galitskii-Migdal formula for the potential 
energy \cite{Gal} ($C_+$ means the closing of the contour in the upper complex $\omega$-plane),
\begin{eqnarray}\label{a2}
v=\frac{1}{2}\int d(k^3)\int\limits_{C_+}\frac{d\omega}{2\pi{\rm i}}\;{\rm e}^{{\rm i}\omega\delta}G(k,\omega)
\Sigma(k,\omega)\ , \quad \delta{_> \atop ^{\to}}0\ , 
\end{eqnarray}
which is related to the total energy $e$ through the virial theorem \cite{Mar}
\begin{eqnarray}\label{a3}
v=r_s\frac{d}{d r_s}e\ . 
\end{eqnarray}
(iii) Besides, $\Sigma(k,\omega)$ appears in the Luttinger theorem ${\rm Im}\ \Sigma(1,\mu)=0$ \cite{Lu}, 
in the Hugenholtz-van Hove theorem $\mu-\mu_0=\Sigma(1,\mu)$ \cite{Hug},  
and in the 
Luttinger-Ward formula for the quasi-particle weight $z_{\rm F}$ \cite{LuWa}:
\begin{equation}\label{a4}
z_{\rm F}=\frac{1}{1-\Sigma'}\ ,\quad \Sigma'=
\left.{\rm Re}\ \frac{\partial\Sigma(1,\omega)}{\partial \omega}\right|_{\omega=\mu}\ .                 
\end{equation} 
So $\Sigma(1,\omega)$ is related to the chemical potential $\mu$, which can be 
calculated from $e$ according to the Seitz theorem \cite{Sei}
\begin{equation}\label{a5}
\mu =\left(\frac{5}{3}-\frac{1}{3}r_s\frac{d}{dr_s}\right)\ e\ ,
\end{equation} 
supposed $e$ is known as a function of $r_s$. Thus from Eq. (\ref{a1}) it follows for $\mu$:
\begin{eqnarray}\label{a6}
\quad \mu_0=\frac{1}{2}, \quad  \mu_{\rm x}=-\frac{\alpha r_s}{\pi}, 
\quad \mu_{\rm c}=(\alpha r_s)^2\left[a\ln r_s+\left(-\frac{1}{3}a+b+b_{2{\rm x}}\right)+O(r_s)\right]\ .
\end{eqnarray}
Similarly as in Eqs. (\ref{a1}) and (\ref{a6}) it is
\begin{equation}\label{a7}
\Sigma(k,\omega)=\Sigma_{\rm x}(k)+\Sigma_{\rm c}(k,\omega), \quad \Sigma_{\rm x}(k)
=-\left(1+\frac{1-k^2}{2k}
\ln\left|\frac{k+1}{k-1}\right|\right)\frac{\alpha r_s}{\pi}\ .
\end{equation}
Notice that $\Sigma(k,\omega)$ in lowest order (of exchange) does not depend on $\omega$. In particular, it is 
$\Sigma_{\rm x}(1)=-\alpha r_s/\pi$, thus $\mu_{\rm x}=
\Sigma_{\rm x}(1)$. Similarly, in 2nd order of exchange the sum rule $\mu_{2{\rm x}}=
\Sigma_{2{\rm x}}(1,\frac{1}{2})$ holds \cite{Zie4}. With 
$\Sigma_{2{\rm x}}(1,\frac{1}{2})=(\alpha r_s)^2 c_{2{\rm x}}$ it takes the form $b_{2{\rm x}}=c_{2{\rm x}}$. 
The  asymptotic behavior $\Sigma_{\rm c}(1,\mu)=(\alpha r_s)^2[a\ln r_s+c+c_{2{\rm x}}+O(r_s)]$ and the sum rule  
\begin{eqnarray}\label{a8}
-\frac{1}{3}a+b=c
\end{eqnarray}
are a must as a consequence of the Hugenholtz-van Hove theorem. But the question is: which partial summation of 
Feynman diagrams has to be used for $\Sigma_{\rm c}$ and what has to be used for $\mu$ ? The obvious answer to the 
first question seems to be $\Sigma_{\rm c}=\Sigma_{\rm r}+ \cdots$ (the subscript "r" means "ring diagram"). 
Symbolically written it is defined as $\Sigma_{\rm r}=G_0\cdot (v_{\rm r}-v_0)$ in terms of the Feynman-diagram 
building elements $G_0$, $v_0$,  and $Q(k,\omega)$ [ = polarization propagator in RPA], cf. Eqs. (\ref{A2}), 
(\ref{A4}), and Fig. {\ref{fey1}}.  $v_{\rm r}$ is the effectively screened Coulomb repulsion following from 
$v_{\rm r}=v_0+v_0Qv_{\rm r}$, see Fig. \ref{fey2}.  The Feynman diagrams of $\Sigma_{\rm r}$ are shown in Fig. 
\ref{fey3}. To what extent this (naive) ansatz has to be changed in a particular way (to answer also the second 
question) will be discussed at the end of Sec. III. \\ 

\noindent
Naively one should expect that in the weak-correlation limit the Coulomb repulsion $\epsilon^2/r$ \cite{foo} 
can be treated as perturbation. But in the early theory of the HEG, Heisenberg \cite{Hei} has shown, that ordinary
perturbation theory with $e_{\rm c}=e_2+e_3+\cdots$ and $e_n\sim (\alpha r_s)^n$ does not apply. Namely, in 2nd order,
there is a direct term $e_{2{\rm d}}$ and an exchange term $e_{2{\rm x}}$, so that $e_2=e_{2{\rm d}}+e_{2{\rm x}}$. 
Whereas $e_{2{\rm x}}/(\alpha r_s)^2$, cf. Fig. \ref{fey4}, 
is a pure finite number $b_{2{\rm x}}$
(not depending on $r_s$), the direct term $e_{2{\rm d}}$ logarithmically diverges along the Fermi surface (i.e. for 
vanishing transition momenta $q$): $e_{2{\rm d}}\to\ln q$ for $q\to 0$. This failure of perturbation theory has been
repaired by Macke \cite{Ma} with an appropriate partial summation of higher-order terms $e_{3{\rm r}},e_{4{\rm r}},
\cdots$ (the subscript "r" means "ring diagram") up to infinite order. This procedure replaces the logarithmic 
divergence for $q\to 0$ by another logarithmic 
divergence, namely for $r_s\to 0$, cf. Eq. (\ref{a1}). This simultaneously "explains", why perturbation theory fails.
The coefficient $a$, first found by Macke \cite{Ma}, has been confirmed later by Gell-Mann and Brueckner
\cite{GB}, who in addition to the logarithmic term numerically calculated two contributions to the next  
term $b$, namely $b_{\rm r}$ and $b_{2{\rm d}}$. More precisely, instead of 
$e_{\rm r}=e_{2{\rm d}}+e_{3{\rm r}}+\cdots$ (notice that $e_{2{\rm r}}=e_{2{\rm d}}$) they calculated a more easily 
doable approximation $e_{\rm r}^0=e_{2{\rm d}}^0+e_{3{\rm r}}^0+\cdots$ (which is sufficient in the weak-correlation 
limit) with the result $e_{\rm r}^0=(\alpha r_s)^2[a \ln r_s +b_{\rm r}+O(r_s)]$, so that
$e_{\rm r}=e_{\rm r}^0+\Delta e_{2{\rm d}}+O(r_s^3)$ with $\Delta e_{2{\rm d}}=e_{2{\rm d}}-e_{2{\rm d}}^0=
(\alpha r_s)^2b_{2{\rm d}}+O(r_s^3)$. In summary,
\begin{eqnarray}\label{a9} 
e_{\rm c}&=&e_{\rm r}+e_{2{\rm x}}+O(r_s^3) \nonumber \\
&=&(e_{2{\rm d}}+e_{3{\rm r}}+\cdots)+e_{2{\rm x}}+O(r_s^3) \nonumber \\
&=&(e_{2{\rm d}}^0+e_{3{\rm r}}^0+\cdots)+(e_{2{\rm d}}-e_{2{\rm d}}^0)+e_{2{\rm x}}+O(r_s^3) \nonumber \\
&=&e_{\rm r}^0+\Delta e_{2{\rm d}}+e_{2{\rm x}}+O(r_s^3) \nonumber \\
&=&(\alpha r_s)^2\{[a\ln r_s+b_{\rm r}+O(r_s)]+b_{2{\rm d}}+b_{2{\rm x}}+O(r_s)\}\ . 
\end{eqnarray}
The result is Eq. (\ref{a1}) with $b=b_{\rm r}+b_{2{\rm d}}$.
This procedure is revisited in Sec. 2 and then in Sec. 3 
applied {\it mutatis mutandis} to the on-the-chemical-potential-shell self-energy $\Sigma(1,\mu)$, the rhs of 
the Hugenholtz-van Hove theorem.
This is a contribution to the mathematics of the weakly-correlated (high-density) HEG). It concerns the HEG 
self-energy in RPA, extending and completing the paper \cite{Cio}. 

\section{The total energy} 
\setcounter{equation}{0}
\noindent
The Heisenberg-Macke story starts with the 2nd-order perturbation theory, $e_2=e_{2{\rm d}}+e_{2{\rm x}}$. Its 
components are the direct ({\rm d}) term $e_{2{\rm d}}$ (with $q_0{_> \atop ^{\rightarrow}}0$) and the exchange 
({\rm x}) term $e_{2{\rm x}}$: 
\begin{equation}\label{b1}
e_{2{\rm d}}=-(\alpha r_s)^2
\frac{2\cdot 3}{(2\pi)^5} \int\limits_{q>q_0}\frac{d^3q\ d^3k_1\ d^3k_2}{ q^4}\; \frac{P}{{\mbox{\bm $q$}} 
\cdot ({\mbox{\bm $k$}}_1+{\mbox{\bm $k$}}_2+{\mbox{\bm $q$}}) }\ , \; k_{1,2}<1, \;
|{\mbox{\bm $k$}}_{1,2}+{\mbox{\bm $q$}}|>1\ ,
\end{equation}
\begin{equation}\label{b2}
e_{2{\rm x}}=+(\alpha r_s)^2
\frac{3}{(2\pi)^5}\int\frac{d^3q\ d^3k_1\ d^3k_2}{ q^2\ ({\mbox{\bm $k$}}_1+{\mbox{\bm $k$}}_2+{\mbox{\bm $q$}})^2}\; 
\frac{P}{{\mbox{\bm $q$}} \cdot ({\mbox{\bm $k$}}_1+{\mbox{\bm $k$}}_2+{\mbox{\bm $q$}}) }\ , \; k_{1,2}<1,\;
|{\mbox{\bm $k$}}_{1,2}+{\mbox{\bm $q$}}|>1\ .
\end {equation}
$P$ means the Cauchy principle value. (Notice the prefactor $-1/2$ and the replacement $q^4\to q^2({\mbox{\bm $k$}}_1+
{\mbox{\bm $k$}}_2+{\mbox{\bm $q$}})^2$, when going from $e_{2{\rm d}}$ to $e_{2{\rm x}}$, and note that the 
2nd-order vacuum diagram of Fig. \ref{fey5} does not contribute.) As already mentioned, the 
integral (\ref{b2}) has been ingeniously calculated by Onsager et al. \cite{Ons} with the result   
$e_{2{\rm x}}=(\alpha r_s)^2\ b_{2{\rm x}}$, $b_{2{\rm x}}=\frac{1}{6}\ln 2-\frac{3}{4}\frac{\zeta(3)}{\pi^2}\approx 
+0.0242$. Unlike $e_{2{\rm x}}$, the direct term $e_{2{\rm d}}$ logarithmically diverges for $q_0\to 0$, i.e. along 
the Fermi surface. This is seen from
\begin{equation}\label{b3}
e_{2{\rm d}}=-(\alpha r_s)^2\frac{2\cdot 3}{(2\pi)^5}\int\limits_{q>q_0} \frac{d^3q}{q^4}I(q)\ ,
\end{equation}
where the Pauli principle makes the function 
\begin{equation}\label{b4}
I(q)= \int\frac{d^3k_1\ d^3k_2\ P}{{\mbox{\bm $q$}} \cdot ({\mbox{\bm $k$}}_1+
{\mbox{\bm $k$}}_2+{\mbox{\bm $q$}})}, \quad k_{1,2}<1, \quad
|{\mbox{\bm $k$}}_{1,2}+{\mbox{\bm $q$}}|>1
\end{equation}
to linearly behave as $I(q\to 0)=\D\frac{8\pi^4}{3}a\ q+O(q^3)$, see App. C, Eq. (\ref{C2}). Thus 
$e_{2{\rm d}}=(\alpha r_s)^2 [a\ln q_0^2+\cdots]$, what agrees with (\ref{a1}) for $q_0^2\sim r_s$. 
The ring-diagram (or RPA) partial summation of Macke \cite{Ma} and Gell-Mann/Brueckner \cite{GB} 
replaces the artificial (by hand) cut-off $q_0$ by a natural cut-off $q_c\sim \sqrt r_s$. This is made  
replacing the divergent direct term $e_{2{\rm d}}$ by the non-divergent ring-diagram sum
\begin{equation}\label{b5}
e_{\rm r}=-\frac{3}{16\pi}\int d^3q\int \frac{d\eta}{2\pi{\rm i}}\ 
\sum\limits_{n=2}^{\infty}\frac{(-1)^n}{n}\left[\left(\frac{q_c}{q}\right)^2Q(q,\eta)\right]^n, \quad 
q_c=\sqrt\frac{4\alpha r_s}{\pi}\ .
\end{equation} 
For $Q(q,\eta)$, the polarization function in lowest order, is given in Eq. (\ref{A4}). 
With $\eta={\rm i}q u$ the contour integration along the real axis is turned to the imaginary axis: 
\begin{equation}\label{b6}
e_{\rm r}=\frac{3}{8\pi}\int du \int\limits_0^\infty d(q^2)
\left[q^2\ln\left(1+\frac{q_c^2}{q^2}\right)R(q,u)-q_c^2R(q,u)\right]\ .
\end{equation}
This has the advantage, that
$R(q,u)=Q(q,{\rm i}qu)$ is a real function, being symmetric in $u$, cf. Eq. (\ref{A2}).
Let us control Eq. (\ref{b6}): The small-$r_s$ expansion of the $u$-integrand starts with $(-1/2)(q_c/q)^4R^2(q,u)$, which just reproduces 
the 2nd-order direct term $e_{2{\rm d}}$ with the help of the integral identity (\ref{C7}). For $r_s\to 0$, a direct 
numerical investigation of Eq. (\ref{b6}) yields $e_{\rm r}\to (\alpha r_s)^2( 0.031091 \ln r_s-0.0711+\cdots )$.
This result is analytically rederived in the following. \\

\noindent
Namely, in the weak-correlation limit $r_s\to 0$ one can approximate $R(q,u)\approx \Theta(q_1-q)R_0(u)+\cdots$ with 
$R_0(u)=1-u\arctan 1/u$, so that $e_{\rm r}=e_{\rm r}^0+O(r_s^3)$, where $e_{\rm r}^0$ contains only the 
$q$-independent $R_0(u)$ and its $q$-integration is restricted to $0\leq q\leq q_1$: 
\begin{eqnarray}\label{b7}
e_{\rm r}^0
&=&\frac{3 }{8\pi}\int\limits_0^\infty du\; \int\limits_0^{q_1} dq^2
\left[q^2\ln[q^2+q_c^2R_0(u)]-q^2\ln q^2-q_c^2R_0(u)\right]  \nonumber \\
&=&\frac{3}{8\pi}\int\limits_0^\infty du\; \frac{1}{2} q_c^4 R_0^2(u)\left[\ln\left(\frac{q_c^2}{q_1^2}R_0(u)\right)-
\frac{1}{2}\right]+O(r_s^3).
\end{eqnarray}
(For a discussion of the divergent/convergent behavior of the $q$-series cf. \cite{GB}, text after their Eq. (23).) 
With $q_c^2=4\alpha r_s/\pi$ it is  
\begin{equation}\label{b8} 
e_{\rm r}^0=(\alpha r_s)^2\frac{3}{\pi^3}\int\limits_0^\infty du\; R_0^2(u)
\left[\ln r_s+\ln \frac{4\alpha}{\pi}-\frac{1}{2}+\ln R_0(u)-2 \ln q_1\right]+O(r_s^3). 
\end{equation}  
So it results $e_{\rm r}^0=(\alpha r_s)^2[a\ln r_s+b_{\rm r}-2a\ln q_1+O(r_s^3)]$ with  
\begin{eqnarray}\label{b9} 
\frac{3}{\pi^3}\int\limits_0^\infty du\; R_0^2(u)=a\approx 0.031091\ , \nonumber \\ 
b_{\rm r}=a\left(\ln \frac{4\alpha}{\pi}-\frac{1}{2}\right)+\frac{3}{\pi^3}\int\limits_0^\infty du\;R_0^2(u)\ln R_0(u)\approx -0.045423\ . 
\end{eqnarray}
For the integrals cf. Eq. (\ref{B3}). \\

\noindent
As it has been explained before and in Eq. (\ref{a9}), the difference between the correct 2nd-order term  of Eq. (\ref{b3}) and 
the first term in the expansion of $e_{\rm r}^0$, namely 
\begin{equation}\label{b10}
e_{2{\rm d}}^0=-(\alpha r_s)^2\ \frac{2\cdot 3}{\pi^3}\int\limits_{q_0}^{q_1}\frac{dq}{q}\int\limits_0^\infty du\; R_0^2(u)+O(r_s^3)
=-(\alpha r_s)^2\ 2a\int\limits_{q_0}^{q_1}\frac{dq}{q}+O(r_s^3)\ , 
\end{equation}
gives 
\begin{eqnarray}\label{b11}
\Delta e_{2{\rm d}}=e_{2{\rm d}}-e_{2{\rm d}}^0&=&-(\alpha r_s)^2\ \frac{2\cdot 3}{\pi^3}\left[\int\limits_{q_0}^\infty \frac{dq}{q}\frac{I(q)}{8\pi q}
-\left(\int\limits_{q_0}^1+\int\limits_1^{q_1}\right)\frac{dq}{q}\frac{\pi^3}{3}a\right]+O(r_s^3)   \\
&=&(\alpha r_s)^2\left\{-\frac{3}{4\pi^4}\left[\int\limits_{q_0}^1 \frac{dq}{q}\left[\frac{I(q)}{q}-\frac{8\pi^4}{3}a\right]+ 
\int\limits_1^\infty \frac{dq}{q}\frac{I(q)}{q}\right]+2a\int\limits_1^{q_1}\frac{dq}{q}\right\}+O(r_s^3)\ . \nonumber 
\end{eqnarray}
$\Delta e_{2{\rm d}}=(\alpha r_s)^2[b_{2{\rm d}}+2a\ln q_1+O(r_s)]$ shows, that the sum $e_{\rm r}^0+\Delta e_{2{\rm d}}$ does not depend on 
$q_1$ for $r_s\to 0$. Besides, the first term of $b_{2{\rm d}}$ is no longer divergent with $q_0\to 0$, therefore it can be set $q_0=0$:
\begin{eqnarray}\label{b12}
b_{2{\rm d}}&=&-\frac{3}{4\pi^4}\int\limits_0^1 \frac{dq}{q}\left[\frac{I(q)}{q}-\frac{8\pi^4}{3}a\
\Theta(1-q)\right] \nonumber \\
&=& \frac{1}{4}+\frac{1}{\pi^2}\left[-\frac{11}{6}-\frac{8}{3}\ln2+2(\ln2)^2\right]\approx -0.025677\ .
\end{eqnarray}
Together it is $b=b_{\rm r}+b_{2{\rm d}}\approx -0.0711$, what agrees with  the above mentioned numerical
evaluation of Eq. (\ref{b6}).

\section{The self-energy}
\setcounter{equation}{0}

\noindent
Here - after the training of Sec. II - , it is aimed to calculate $\Sigma_{\rm c}(1,\mu)$ in the weak-correlation 
limit, 
where there is a scheme for $\Sigma_{\rm c}(1,\mu)$ analog to Eq. (\ref{a9}) for $e_{\rm c}$ with one difference. 
Namely, whereas the chemical-potential shift $\mu$ results from vacuum diagrams, the self-energy $\Sigma(k,\omega)$ 
results from non-vacuum diagrams, which are functions of $k$ and $\omega$, see the discussion at the end of this 
Section. \\ 

\begin{figure}[h!]
\begin{center}
\rotatebox{0}{\resizebox{100mm}{!}{\includegraphics{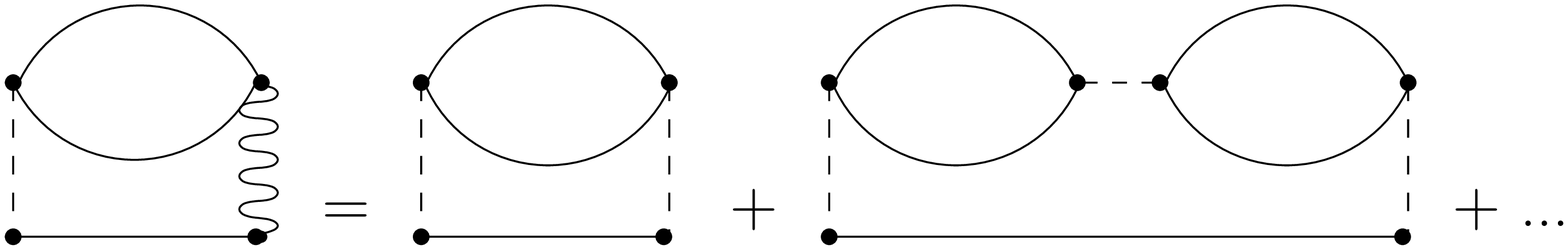}}}
\caption{\label{fey3}Feynman diagrams for the ring-diagram-summed self-energy 
$\Sigma_{\rm r}=G_0\cdot(v_{\rm r}-v_0)$ as defined in Eq. (\ref{c4}).}
\end{center}
\end{figure}

\begin{figure}[h!]
\begin{center}
\rotatebox{0}{\resizebox{75mm}{!}{\includegraphics{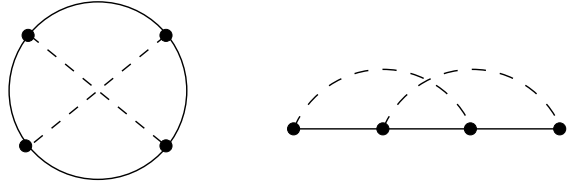}}}
\caption{\label{fey4}Feynman diagrams for $e_{2{\rm x}}$ \cite{Ons} and $\Sigma_{2{\rm x}}$ \cite{Zie4}.}
\end{center}                                                                                     
\end{figure}

\begin{figure}[h!]
\begin{center}
\rotatebox{0}{\resizebox{75mm}{!}{\includegraphics{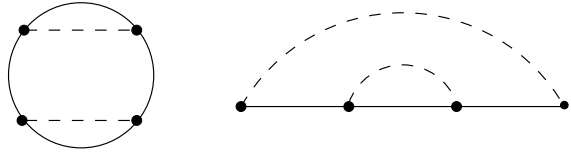}}}
\caption{\label{fey5}Feynman diagrams, which do not contribute to $e_2$ and $\Sigma_2$, respectively.}
\end{center}
\end{figure}

\noindent
In analogy to Eqs. (\ref{b1}) and (\ref{b2}), the self-energy in 2nd order is $\Sigma_2(k,\omega)
=\Sigma_{2{\rm d}}(k,\omega)+\Sigma_{2{\rm x}}(k,\omega)$, the 2nd-order self-energy diagram of Fig. \ref{fey5}
vanishes. From (\ref{A6}) it follows for the direct term
\begin{eqnarray}\label{c1}
\Sigma_{2{\rm d}}(1,\omega)=\frac{(\alpha r_s)^2}{2\pi^4} \int\limits_{q>q_0}\frac{d^3qd^3k_2}{q^4}
&&\left[\frac{\Theta(|{\mbox{\bm $e$}}_1+{\mbox{\bm $q$}}|-1)}{\omega-\frac{1}{2}-{\mbox{\bm $q$}} \cdot 
({\mbox{\bm $e$}}_1+{\mbox{\bm $k$}}_2+{\mbox{\bm $q$}})+{\rm i}\delta}\right.  \\
&&+\left.\frac{\Theta(1-|{\mbox{\bm $e$}}_1+
{\mbox{\bm $q$}}|)}{\omega-\frac{1}{2}-{\mbox{\bm $q$}}\cdot({\mbox{\bm $e$}}_1-{\mbox{\bm $k$}}_2)-{\rm i}\delta}\right]\Theta(1-k_2)
\Theta(|{\mbox{\bm $k$}}_2+{\mbox{\bm $q$}}|-1). \nonumber 
\end{eqnarray}
For the corresponding exchange term $\Sigma_{2{\rm x}}(1,\omega)$ cf. Fig. \ref{fey4} and ref. \cite{Zie4}, where it has been shown that
$\Sigma_{2{\rm x}}={\rm Re}\ \Sigma_{2{\rm x}}(1,\frac{1}{2})=(\alpha r_s)^2c_{2{\rm x}}$ with the sum rule $c_{2{\rm x}}=b_{2{\rm x}}
\approx +0.0242$. On the other hand, the direct term $\Sigma_{2{\rm d}}$ diverges logarithmically for 
$q_0\to 0$. This is seen from 
\begin{equation}\label{c2}
\Sigma_{2{\rm d}}={\rm Re}\ \Sigma_{2{\rm d}}\left(1,\frac{1}{2}\right)= -\frac{(\alpha r_s)^2}{2\pi^4}\int\limits_{q>q_0} \frac{d^3q}{q^4}J(q)\ ,
\end{equation}
where the Pauli principle makes the function
\begin{eqnarray}\label{c3} 
J(q)=\int \frac{d^2e_1}{4\pi}d^3k_2
\left[\frac{\Theta(|{\mbox{\bm $e$}}_1+{\mbox{\bm $q$}}|-1)P}{{\mbox{\bm $q$}}\cdot
({\mbox{\bm $e$}}_1+{\mbox{\bm $k$}}_2+{\mbox{\bm $q$}})}+
\frac{\Theta(1-|{\mbox{\bm $e$}}_1+{\mbox{\bm $q$}}|)P}{{\mbox{\bm $q$}}\cdot
({\mbox{\bm $e$}}_1-{\mbox{\bm $k$}}_2)}\right]
\Theta(1-k_2)\Theta(|{\mbox{\bm $k$}}_2+{\mbox{\bm $q$}}|-1)\ , \nonumber \\
\end{eqnarray}
to linearly behave as $J(q\to 0)=\pi^3a\ q+O(q^3)$, see App. D, Eq. (\ref{D2}). Thus 
$\Sigma_{2{\rm d}}=(\alpha r_s)^2(a\ln q_0^2+\cdots)$, what is for $q_0^2\sim r_s$ in full agreement with the 
Hugenholtz-van Hove theorem (\ref{a4}) and the perturbation expansion of $\mu$, which - because of (\ref{a5}) - 
gives $\mu_{2{\rm d}}=e_{2{\rm d}}=(\alpha r_s)^2(a\ln r_s+\cdots)$. In the ring-diagram partial summation the 
divergent direct term $\Sigma_{2{\rm d}}(k,\omega)$ is replaced by the non-divergent sum (its Feynman diagrams 
are shown in Fig. \ref{fey3})
\begin{eqnarray}\label{c4}
\Sigma_{\rm r}(k,\omega)&=&(\alpha r_s)^2\ \frac{2}{\pi^3}\int \frac{d^3q}{q^2}\int \frac{d\eta}{2\pi{\rm i}}\
\frac{Q(q,\eta)}{q^2+q_c^2Q(q,\eta)}\times \nonumber \\
&\times&\left[\D\frac{\Theta(|{\vek k}+{\mbox{\bm $q$}}|-1)}
{\omega+\eta-\frac{1}{2}k^2-{\mbox{\bm $q$}}\cdot(\vek k+\frac{1}{2}\mbox{\bm $q$})+{\rm i}\delta}+
\D
\frac{\Theta(1-|{\vek k}+{\mbox{\bm $q$}}|)}{\omega+\eta-\frac{1}{2}k^2-{\mbox{\bm $q$}}\cdot(\vek k+\frac{1}{2}\mbox{\bm $q$})-{\rm i}\delta}\right] . 
\end{eqnarray}
Next, this expression is carefully controlled:\\
(i) If the term $q_c^2Q(q,\eta)$, which describes the RPA screening of the bare Coulomb interaction $1/q^2$, is 
deleted, then $\Sigma_{\rm r}(k,\omega)$ changes to $\Sigma_{2\rm d}(k,\omega)$, as it is seen from Eq. (\ref{A5}). \\
(ii) Use of Eq. (\ref{c4}) in the Galitskii-Migdal formula (\ref{a2}) yields the ring-diagram summation for the 
potential energy, $v_{\rm r}$, which follows from $e_{\rm r}$ through the virial theorem (\ref{a3}). \\ 
(iii) The expression (\ref{c4}) allows to calculate the derivative $\Sigma_{\rm r}'(k,\omega)=
\partial\Sigma_{\rm r}(k,\omega)/\partial \omega$. From $\Sigma_{\rm r}'={\rm Re}\ \Sigma_{\rm r}'(1,\frac{1}{2})$ 
one obtains $z_{\rm F}$ in RPA by means of the Luttinger-Ward formula (\ref{a4}) as 
$z_{\rm F}=1+\Sigma_{\rm r}'+\cdots$ with 
\begin{eqnarray}\label{c5}
\Sigma_{\rm r}'=\frac{\alpha r_s}{\pi^2}\int\limits_0^\infty du\ \frac{R_0'(u)}{R_0(u)}\arctan\frac{1}{u}+\cdots=
-0.177038\ r_s+\cdots\ . 
\end{eqnarray}
This is just the well-known RPA result for $z_{\rm F}$ \cite{Da}. For the integral see Eq. (\ref{B4}). 
 \\

\begin{figure}[h!]
\begin{center}
\rotatebox{0}{\resizebox{100mm}{!}{\includegraphics{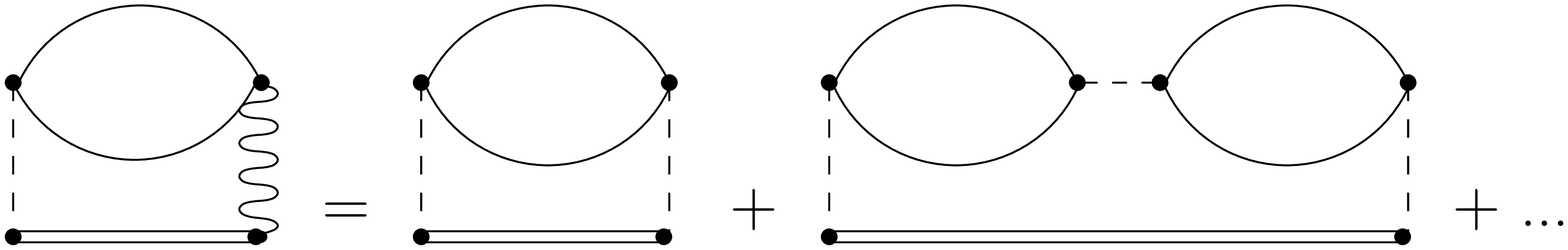}}}
\caption{\label{fey6}Feynman diagrams for $\Sigma_{\rm r}^{\rm x}=G_{\rm x}\cdot(v_{\rm r}-v_0)$. For $G_{\rm x}$
see Fig. \ref{fey7} and Eq. (\ref{c14})} 
\end{center}
\end{figure}

\begin{figure}[h!]
\begin{center}
\rotatebox{0}{\resizebox{100mm}{!}{\includegraphics{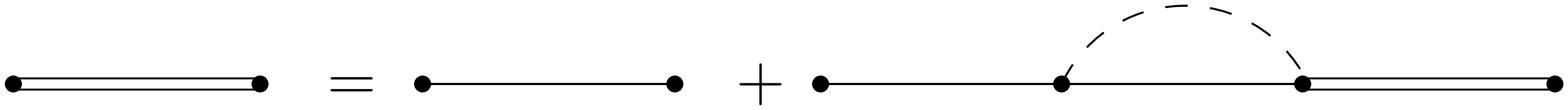}}}
\caption{\label{fey7}Feynman diagrams for the renormalized one-body Green's function 
$G_{\rm x}=G_0+G_0\Sigma_{\rm x}G_{\rm x}$, see Eq. (\ref{c14}).}
\end{center}
\end{figure}

\noindent
After this controlling and training, $\Sigma_{\rm r}={\rm Re}\ \Sigma_{\rm r}(1,\frac{1}{2})$ is derived from Eq. 
(\ref{c4}) in a similar way as $e_{\rm r}$ in Eqs. (\ref{b7}) - (\ref{b12}).
The next steps again are  substitution $\eta={\rm i}qu$ and contour deformation from the real to the imaginary 
axis with $x={\vek e}\cdot{\vek e}_q$ and $|{\mbox {\bm $e$}}+{\mbox {\bm $q$}}|{_> \atop ^{<}}1\pm\delta$:  
\begin{eqnarray}\label{c6}
\Sigma_{\rm r}
&=&-\frac{(\alpha r_s)^2}{\pi^4}\int\frac{d^3q}{q^2}\int du\ \frac{R(q,u)}{q^2+q_c^2R(q,u)}\cdot 
\frac{1}{(x+\frac{q}{2})-{\rm i}u} \nonumber \\
&=&-\frac{(\alpha r_s)^2}{\pi^4}\int\limits_0^\infty du\ \int\frac{d^3q}{q^2}\ \frac{R(q,u)}{q^2+q_c^2R(q,u)}\cdot 
\frac{2(x+\frac{q}{2})}{(x+\frac{q}{2})^2+u^2}\ . 
\end{eqnarray}
In the last step the $u$- and ${\mbox{\bm $q$}}$-integrations are exchanged and 
it is used that $R(q,u)$ is even in $u$, cf. Eq. (\ref{B1}); so the imaginary part again vanishes.  
Next the ${\mbox{\bm $q$}}$-integration is specified as 
\begin{eqnarray}\label{c7}
\Sigma_{\rm r}&=&
-\frac{(\alpha r_s)^2}{\pi^4}\int\limits_0^{\infty}du\int\frac{d^3q}{q^2}\ 
\frac{R(q,u)}{q^2+q_c^2R(q,u)}\cdot
\int\limits_{-1}^{+1}\frac{dx}{2}\ \frac{2(x+\frac{q}{2})}{(x+\frac{q}{2})^2+u^2} \nonumber \\
&=&-(\alpha r_s)^2\frac{2}{\pi^3}\ \int\limits_0^{\infty}du\ \int\limits_0^{\infty}dq\
\frac{R(q,u)}{q^2+q_c^2R(q,u)}\cdot
\ln\frac{(\frac{q}{2}+1)^2+u^2}{(\frac{q}{2}-1)^2+u^2}\ .
\end{eqnarray}
Let us control Eq. (\ref{c7}): The small-$r_s$ expansion of the $u$-integrand starts with $R(q,u)/q^2$, which just
reproduces the 2nd-order direct term (\ref{c3}) with the help of the integral identity (\ref{D7}). 
In the limit $r_s\to 0$, Eq. (\ref{c7}) numerically gives $\Sigma_{\rm r}\approx (\alpha r_s)^2 
(0.031091 \ln r_s -0.081463+\cdots )$. This result is analytically confirmed by the following. 
The asymptotic behavior for $r_s\to 0$ is determined by the lower integration limit $q\to 0$, therefore $R(q,u)$ and 
$\ln \cdots$ can be approximated by $R_0(u)=1-u\arctan 1/u$ and $L_0(u)=2q/(1+u^2)$, respectively:
\begin{equation}\label{c8}
\Sigma_{\rm r}=-(\alpha r_s)^2\frac{2}{\pi^3}\int\limits_0^{\infty}du\int\limits_0^\infty d(q^2)\
\frac{[R_0(u)+R_1(u)q^2+\cdots][L_0(u)+L_1(u)q^2+\cdots]}{q^2+q_c^2[R_0(u)+R_1(u)q^2+\cdots]}
=\Sigma_{\rm r}^0+O(r_s^3)\ .
\end{equation} 
Finally the $q$-integration yields
\begin{eqnarray}\label{c9}
\Sigma_{\rm r}^0
&=&-(\alpha r_s)^2\frac{2}{\pi^3}\int\limits_0^\infty du\; \frac{R_0(u)}{1+u^2}\int\limits_0^{q_2}
\frac{dq^2}{q^2+q_c^2R_0(u)} \nonumber \\
&=& -(\alpha r_s)^2\frac{2}{\pi^3}\int\limits_0^{\infty}du\; \frac{R_0(u)}{1+u^2}
\left.\ln[q^2+q_c^2R_0(u)]\right|_0^{q_2} \nonumber \\
&=&-(\alpha r_s)^2\frac{2}{\pi^3}\int\limits_0^\infty du\ \frac{R_0(u)}{1+u^2}\ [2\ln q_2-\ln q_c^2R_0(u)]\ . 
\end{eqnarray}
With Eq. (\ref{B3}) it turns out $\Sigma_{\rm r}^0=(\alpha r_s)^2[ a\ln r_s+c_{\rm r}-2a\ln q_2+O(r_s)]$,
\begin{equation}\label{c10}
c_{\rm r}
= a\ln \frac{4\alpha}{\pi}+\frac{2}{\pi^3}\int\limits_0^\infty du\ \frac{R_0(u)\ln R_0(u)}{1+u^2} \approx -0.035059\ . 
\end{equation}
The difference between the exact 2nd-order term of Eq. (\ref{c2}) and the first term in the $q$-expansion of 
$\Sigma_{\rm r}^0$, namely
\begin{equation}\label{c11}
\Sigma_{2{\rm d}}^0=-(\alpha r_s)^2\frac{2}{\pi^3}\int\limits_{q_0}^{q_2}\frac{dq}{q^2}\int\limits_0^\infty du\ R_0(u)\ 
\frac{2q}{1+u^2}=-(\alpha r_s)^2\frac{2}{\pi^3}\int\limits_{q_0}^{q_2}\frac{dq}{q}\pi(1-\ln 2)\ ,    
\end{equation}
yields
\begin{eqnarray}\label{c12}
\Delta\Sigma_{2{\rm d}}=\Sigma_{2{\rm d}}-\Sigma_{2{\rm d}}^0
&=&-(\alpha r_s)^2\frac{2}{\pi^3}\left[\int\limits_{q_0}^\infty\frac{dq}{q}\frac{J(q)}{q}-
\int\limits_{q_0}^{q_2}\frac{dq}{q}\pi(1-\ln 2)\right] \\ 
&=&(\alpha r_s)^2\left\{-\frac{2}{\pi^3}\; \left [\int\limits_{q_0}^1\frac{dq}{q}\left[\frac{J(q)}{q}-\pi(1-\ln 2)\right]+
\int\limits_1^\infty\frac{dq}{q}\frac{J(q)}{q}\right ] +2a\int\limits_1^{q_2}\frac{dq}{q}\right\}\ . \nonumber 
\end{eqnarray}
$\Delta\Sigma_{2{\rm d}}=(\alpha r_s)^2[c_{2{\rm d}}+2a\ln q_2+\cdots]$ shows, that the sum $\Sigma_{\rm r}^0+\Delta\Sigma_{2{\rm d}}$ does not
depend on $q_2$ for $r_s\to 0$. Besides the first term of $c_{2{\rm d}}$ is no longer divergent with $q_0\to 0$, therefore it can be set 
$q_0=0$:
\begin{equation}\label{c13}
c_{2{\rm d}}=-\frac{2}{\pi^3}\;  \int\limits_0^\infty\frac{dq}{q}\left[\frac{J(q)}{q}-\pi(1-\ln 2)
\Theta(1-q)\right]\approx -0.046404\ .
\end{equation}
For the $J(q)$-integral cf. Eq. (\ref{D4}). Together it is $c=c_{\rm r}+c_{2{\rm d}}\approx 
-0.08146$, to be compared with $-\frac{1}{3}a+b=-0.08146$. This is just the expected sum rule (\ref{a8}). But the 
difference $\Sigma_{\rm r}(1,\frac{1}{2}+\mu_{\rm x}\cdots)-\Sigma_{\rm r}(1,\frac{1}{2})\neq 0 $ seems to disturb 
this sum rule $c=-\frac{1}{3}a+b$. This 'misfit' is removed by an additional partial summation replacing 
Fig. \ref{fey3} by Fig. \ref{fey6}, i.e. replacing 
$\Sigma_{\rm r}=G_0\cdot (v_{\rm r}-v_0)$ by $\Sigma_{\rm r}^{\rm x}=G_{\rm x}\cdot (v_{\rm r}-v_0)$ with the renormalized 
one-body Green's function 
\begin{equation}\label{c14}
G_{\rm x}(k,\omega)=\frac{1}{\omega -\frac{1}{2}k^2-\Sigma_{\rm x}(k)\pm{\rm i}\delta}\ , 
\end{equation}
see Fig. \ref{fey7}. For $k=1$ it is $G_{\rm x}(1,\omega)=
1/[\omega -\frac{1}{2}-\mu_{\rm x}\pm{\rm i}\delta]$. Thus $\Sigma_{\rm r}^{\rm x}(1,\frac{1}{2}+\mu_{\rm x}+\cdots)=
\Sigma_{\rm r}(1,\frac{1}{2})+\cdots$ in the limit $r_s\to 0$. So the conjectured relation (\ref{a8}) holds. This
can be seen still in another way. \\ 

\noindent
Namely, note the similarities of Eqs. (\ref{c10}) and (\ref{b9}) as well as Eqs. (\ref{c13}) and (\ref{b12}). Their 
differences are 
\begin{equation}\label{c15}
c_{\rm r}-b_{\rm r}=\frac{1}{3}a \quad {\rm and} \quad c_{2{\rm d}}-b_{2{\rm d}}= - \frac{2}{3}a 
\end{equation} 
using the identities (\ref{B5}) and (\ref{D8}), respectively. Thus, with $b=b_{\rm r}+b_{2{\rm d}}$ and
$c=c_{\rm r}+c_{2{\rm d}}$, the sum rule (\ref{a8}) is proven once more.  

\section{Summary}
\begin{figure}[h!]
\begin{center}
\rotatebox{0}{\resizebox{100mm}{!}{\includegraphics{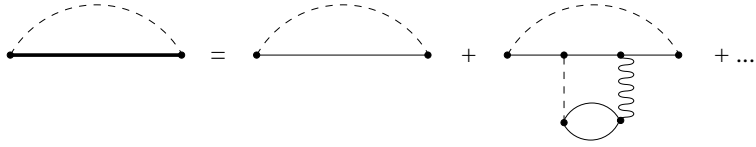}}}
\caption{\label{fey8}Feynman diagrams for $\Sigma^{\rm HF}=Gv_0$ with $G$ = full one-body Green's function
of the interacting system, $G=G_0+G_0\Sigma G$, $\Sigma$ = full self-energy. The lowest-order term is 
$\Sigma_{\rm x}=G_0v_0$, therefore the correlation part is $\Sigma_{\rm c}^{HF}=(G-G_0)v_0$.}
\end{center}
\end{figure}

\setcounter{equation}{0}
\noindent
Summarizing, the Hugenholtz-van Hove theorem $\mu-\mu_0=\Sigma(1,\mu)$ takes for the HEG ground state in its 
weak-correlation limit $r_s\to 0$ the 
asymptotic form 
\begin{eqnarray}\label{d1}
-\frac{\alpha r_s}{\pi}+(\alpha r_s)^2[a \ln r_s +&\left(-\frac{1}{3}a+b_{\rm r}+
b_{2{\rm d}}\right)&+b_{2{\rm x}}+O(r_s)]= \nonumber \\
-\frac{\alpha r_s}{\pi}+(\alpha r_s)^2[a \ln r_s+&(c_{\rm r}+c_{2{\rm d}})&+c_{2{\rm x}}+O(r_s)]\ .
\end{eqnarray} 
So the sum rules [with $a=\frac{1}{\pi^2}(1-\ln 2)$ and $b_{{\rm r}}, b_{2{\rm d}}, c_{\rm r}, c_{2{\rm d}}$ given
in Eqs. (\ref{b9}), (\ref{b12}), (\ref{c10}), (\ref{c13})] 
\begin{equation}\label{d2}
\frac{1}{3}a+b_{\rm r}=c_{\rm r}, \quad -\frac{2}{3}a+b_{2{\rm d}}= c_{2{\rm d}}, \quad b_{2{\rm x}}=c_{2{\rm x}}
\end{equation}
hold. The last relation or $\mu_{2{\rm x}}=\Sigma_{2{\rm x}}$ has been 
shown in \cite{Zie4}. The sum rules (\ref{d2}) are relations between the Macke number $a$, the Gell-Mann/Brueckner 
numbers $b_{\rm r}$, $b_{2{\rm d}}$ and the Onsager/Mittag/Stephen number $b_{2{\rm x}}$ (which altogether desribe 
the 
$r_s\to 0$ asymptotics of the correlation energy $e_{\rm c}$) on the one hand and corresponding numbers $c_{\rm r}$,
$c_{2{\rm d}}$, $c_{2{\rm x}}$ of the on-shell self-energy $\Sigma (1,\mu)$ on the other hand. 
Eqs. (\ref{d1}) and (\ref{d2}) answer the question which partial summation of Feynman diagrams 
has to be used in the weak-correlation limit for the self-energy $\Sigma$ on the rhs of the Hugenholtz-van Hove 
theorem. They 
result from the renormalized ring-diagram (or RPA) partial summation (symbolically written as)
$\Sigma\approx \Sigma_{\rm r}^{\rm x}=G_{\rm x}\cdot v_{\rm r}$ with $v_{\rm r}=v_0/(1-Qv_0)$ and $G_{\rm x}(k,\omega)$ = renormalized one-body Green's
function and $Q(q,\eta)$ = polarization propagator in RPA, see Eqs. (\ref{c14}) and (\ref{A4}), respectively, and Figs. \ref{fey1}, \ref{fey2}, 
\ref{fey6}, \ref{fey7}. (They not result from $\Sigma\approx \Sigma^{\rm HF}= G\cdot v_0$, see Fig. \ref{fey8},  
as an alternative ansatz with
$G(k,\omega)$ = full one-body Green's function of the interacting system \cite{foo2}.) Byproducts are the analytical 
representation of $b_{2{\rm d}}$, a detailed description of the momentum-transfer or Macke function $I(q)$ for the
RPA vacuum diagrams (App. C), the introduction and discussion of an analog function $J(q)$ for the RPA 
self-energy diagrams (App. D), and the proof of integral identities, which relate $I(q)$ and $J(q)$ to the 
polarization-propagator function $R(q,u)$, cf. Apps. C and D. 

\section*{Acknowledgments}
\noindent
The author thanks G. Diener and E. Runge for valuable hints  
and F. Tasn\'adi for technical help and acknowledges P. Fulde for supporting this work.

\begin{appendix}
\section*{Appendix A: One-body Green's function, particle-hole propagator, and 2nd-order self-energy}
\setcounter{equation}{0}
\renewcommand{\theequation}{A.\arabic{equation}}
\noindent
In the following the identities (with $z=x+{\rm i}y$)
\begin{equation}
\label{A1}
 \int\limits_{C_\pm}\frac{dz}{2\pi{\rm i}}\frac{1}{(z-z_1)(z-z_2)}= 
 \left\{\begin{array}{ccl}
    0~                   && {\rm for}\;{\rm sign}\ y_1={\rm sign}\ y_2  \\[+5mm]
    {\D\frac{1}{z_1-z_2}~}
                            && {\rm for}\; y_1>0\; {\rm and}\; y_2<0           \\[+5mm]
    {\D\frac{1}{z_2-z_1}~}
                            && {\rm for}\; y_1<0\; {\rm and}\; y_2>0                 \\
    
  \end{array}
 \right.
\end{equation}
for contour integrations in the complex $z$-plane are used ($z=x+{\rm i}y$, $C_{\pm}$= contour along 
the real axis, closed above or below with a half circle). The building elements of the Feynman diagrams are 
\begin{equation}\label{A2}
G_0(k,\omega)=\frac{\Theta(k-1)}{\omega-\frac{1}{2}k^2+{\rm i}\delta}+
\frac{\Theta(1-k)}{\omega-\frac{1}{2}k^2-{\rm i}\delta}\ , \quad \delta{_> \atop ^{\to}}0 \quad {\rm and} \quad
v_0(q)=\frac{4\pi\alpha r_s}{q^2}\ . 
\end{equation}
From $G_0$ follows the particle-hole propagator $Q$ in RPA according to
\begin{eqnarray}\label{A3}
Q(q,\eta)=-\int\frac{d^3k}{4\pi}\int\frac{d\omega}{2\pi{\rm i}}G_0(k,\omega)
G_0(|{\mbox{\bm $k$}}+{\mbox{\bm $q$}}|,\omega+\eta)
\end{eqnarray}
with the result
\begin{eqnarray}\label{A4}
Q(q,\eta)=\int \frac{d^3k}{4 \pi} \left[\frac{1}{\mbox{\bm $q$}({\mbox{\bm $k$}}
+\frac{1}{2}{\mbox{\bm $q$}})-\eta-{\rm i}\delta}+
\frac{1}{{\mbox{\bm $q$}}({\mbox{\bm $k$}}+\frac{1}{2}{\mbox{\bm $q$}})+\eta-{\rm i}\delta}\right]
\Theta(1-k)\Theta(|{\mbox{\bm $k$}}+{\mbox{\bm $q$}}|-1). \nonumber \\
\end{eqnarray}
(\ref{A2}) and (\ref{A4}) used in the direct term of the 2nd-order off-shell self-energy
\begin{equation}\label{A5}
\Sigma_{2{\rm d}}(k,\omega)=(\alpha r_s)^2\ \frac{2}{\pi^3}\int\limits_{q>q_0}\frac{d^3q}{q^4}\int\frac{d\eta}{2\pi
{\rm i}} \; Q(q,\eta) G_0(|{\mbox{\bm $k$}}+{\mbox{\bm $q$}}|,\omega+\eta)
\nonumber \\
\end{equation}
yields
\begin{eqnarray}\label{A6}
\Sigma_{2{\rm d}}(k,\omega)=\frac{(\alpha r_s)^2}{2\pi^4} \int\limits_{q>q_0}\frac{d^3q}{q^4}\int d^3k'
&\left[ 
\D\frac{\Theta(|{\mbox{\bm $k$}}+{\mbox{\bm $q$}}|-1)}{\omega-\frac{1}{2}k^2-{\mbox{\bm $q$}} 
\cdot ({\mbox{\bm $k$}}+{\mbox{\bm $k$}}'+{\mbox{\bm $q$}})+{\rm i}\delta}
\right. & \\
&+\left.
\D\frac{\Theta(1-|{\mbox{\bm $k$}}+
{\mbox{\bm $q$}}|)}{\omega-\frac{1}{2}k^2-{\mbox{\bm $q$}}\cdot({\mbox{\bm $k$}}-{\mbox{\bm $k$}}')-{\rm i}\delta}
\right]&\Theta(1-k') \Theta(|{\mbox{\bm $k$}}'+{\mbox{\bm $q$}}|-1).  \nonumber 
\end{eqnarray}
This expression used in (\ref{a2}) yields $v_{2{\rm d}}=2e_{2{\rm d}}$  
in agreement with the virial theorem (\ref{a3}).

\section*{Appendix B: The function $R(q,u)$}
\setcounter{equation}{0}
\renewcommand{\theequation}{B.\arabic{equation}}
\noindent
$Q(q,\eta)$ becomes real for imaginary $\eta$:
\begin{eqnarray}\label{B1}
R(q,u)=Q(q,{\rm i}qu)=\frac{1}{2}
\left[
1+\frac{1+u^2-\frac{q^2}{4}}{2q}
\ln\frac{(\frac{q}{2}+1)^2+u^2}{(\frac{q}{2}-1)^2+u^2} \right. \nonumber \\
\left.  - u\left( \arctan\frac{1+\frac{q}{2}}{u}+\arctan\frac{1-\frac{q}{2}}{u} \right)
\right] .
\end{eqnarray}
The function $R(q,u)$ has the $q$-expansion $R(q,u)=R_0(u)+q^2R_1(u)+\cdots$ with
\begin{eqnarray}\label{B2}
R_0(u)=1-u\arctan \frac{1}{u}\ ,\quad R_1(u)=-\frac{1}{12(1+u^2)^2}\ , \quad R_2(u)=-\frac{1-5u^2}{240(1+u^2)^4}\ . 
\end{eqnarray}
Here is a list of integrals:
\begin{eqnarray}\label{B3}
\int\limits_0^\infty du\; R_0^2(u)&=&\frac{\pi^3}{3}a\approx 0.321336,\quad 
\int\limits_0^\infty du \; R_0^2(u)\ln R_0(u)\approx -0.176945, \nonumber \\ 
\int\limits_0^\infty du\ \frac{R_0(u)}{1+u^2}&=&\frac{\pi^3}{2}a\approx 0.482003, \quad
\int\limits_0^\infty du\ \frac{R_0(u)\ln R_0(u)}{1+u^2}\approx-0.345751, 
\end{eqnarray}
\begin{equation}\label{B4}
\int\limits_0^\infty du\; \frac{R'_0(u)}{R_0(u)}\arctan \frac{1}{u}\approx -3.353337\ , \quad 
\int\limits_0^\infty du\; \frac{R''_0(u)}{R_0(u)}\arctan \frac{1}{u}\approx  4.581817\ .
\end{equation}
The last but one integral appears in the weak-correlation limit of the quasi-particle weight 
$z_{\rm F}$ \cite{Da}. The identity 
\begin{equation}\label{B5}
\frac{2}{\pi^3}\int\limits_0^\infty du\ R_0(u)\ln R_0(u)\left[\frac{1}{1+u^2}-\frac{3}{2}R_0(u)\right]=-\frac{1}{6}a
\end{equation}
leads to the sum rule (\ref{c15}). \\

\section*{Appendix C: The function $I(q)$}
\setcounter{equation}{0}
\renewcommand{\theequation}{C.\arabic{equation}}

\noindent
Using cylindrical coordinates and the centre of the 
vector $\mbox{\bm $q$}$ as origin, Macke \cite{Ma} succeeded to calculate $I(q)$ explicitly as  
\begin{eqnarray}
I(q\leq 2)=
\pi^2\left[\left(\frac{29}{15}\right.\right.-\left.\left.\frac{8}{3}\ln 2\right) q-\frac{q^3}{20}\right.
&+&\frac{1}{q}\left(\frac{16}{15}+q-\frac{q^3}{6}+\frac{q^5}{80}\right)\ln \left(1+\frac{q}{2}\right) \nonumber \\
&+&\left.\frac{1}{q}\left(\frac{16}{15}-q+\frac{q^3}{6}-\frac{q^5}{80}\right)\ln \left(1-\frac{q}{2}\right)\right], 
\nonumber 
\end{eqnarray}
\begin{eqnarray}\label{C1}
I(q\geq 2)=\frac{\pi^2}{30}\left[4\; (22+q^2)\right.
&+&\frac{1}{q}(q+2)^3(4-6q+q^2)\ln \left(1+\frac{2}{q}\right) \nonumber \\
&+&\left.\frac{1}{q}(q-2)^3(4+6q+q^2)\ln \left(1-\frac{2}{q}\right)\right].
\end{eqnarray}
Therefore $I(q)$ is referred to as Macke function, cf. also \cite{Ho}. 
(The last two lines of (\ref{C1}) correct errors in \cite{Zie2}, Eq. (A.1).) $I(q)$ has the properties
\begin{eqnarray}\label{C2}
I(q\to 0)&=&\frac{8 \pi^2}{3}(1-\ln 2)\ q-\frac{\pi^2}{6}\; q^3+\cdots, \quad 
I(q\to \infty)=\left(\frac{4\pi}{3}\right)^2\left( \frac{1}{q^2}+\frac{2}{5}\frac{1}{q^4}+\cdots\right ), \nonumber \\   
I(2)&=&\frac{4\pi^2}{15}(13-16\ln 2)\approx 5.02598, \quad I'(2)=-\frac{8\pi^2}{5}(-1+2\ln 2)\approx -6.10012, \nonumber \\ 
I''(2^+)&=&\frac{8\pi^2}{15}(2+\ln 2)\approx 14.1762, \quad I''(2^-)=-\frac{2\pi^2}{15}(7-4\ln 2)\approx -5.56305.
\end{eqnarray}
$I(q)$ has a maximum of $7.12$ at $q=1.36$. $I(q)$ and $I'(q)$ are continuous at $q=2$, but $I''(q)$ 
has there a jump discontinuity of $2\pi^2$. This is because the topology changes from overlapping to 
non-overlapping Fermi spheres, when passing $q=2$ from below. Its  normalization is  
\begin{equation}\label{C3}
\int_0^\infty dq\; I(q)=\frac{8\pi^2}{45}(-3+\pi^2+6\ln2)\approx 19.3505.
\end{equation}
$I(q)$ is shown in Fig. 9. Multiplying the integral
\begin{eqnarray}\label{C4}
\int\limits_0^\infty \frac{dq}{q^2}\; [I(q)-\frac{8\pi^2}{3}(1-\ln 2)\ q \Theta(1-q)]&=&
\frac{\pi^2}{9}\ [22-3\pi^2+32\ln 2-24(\ln 2)^2]\approx 3.334856 \nonumber \\  
\end{eqnarray}
with $-3/(4\pi^4)$ yields Eq. (\ref{b12}). \\

\noindent
The original expression for $I(q)$ arises from the diagram rules for $e_{2{\rm d}}$ with 
Eq. (\ref{c3}) as 
\begin{equation}\label{C5}
I(q)=\frac{1}{2}(4\pi)^2{\rm Re}\ \int \frac{d\eta}{2\pi{\rm i}}\; Q^2(q,\eta).
\end{equation}
One way is to insert (\ref{A4}) into (\ref{C5}). It results (with $x_i={\mbox{\bm $q$}}\cdot{(\mbox{\bm $k$}}_i+
\frac{1}{2}{\mbox{\bm $q$}}), i=1,2$)
\begin{equation}\label{C6}
I(q)={\rm Re}\ \int \frac{d^3k_1d^3k_2}{x_1+x_2-{\rm i}\delta } 
=\int \frac{d^3k_1d^3k_2\ P}{{\mbox{\bm $q$}}\cdot({\mbox{\bm $k$}}_1+{\mbox{\bm $k$}}_2+{\mbox{\bm $q$}})}
\end{equation} 
in agreement with Eq. (\ref{b4}). Another way is 
the analytical continuation and the deformation of the integration contour from the real 
to the imaginary axis with the advantage that $R(q,u)=Q(q,{\rm i}qu)$ is a real function. This 
yields
\begin{equation}\label{C7}
I(q)=2\cdot 4\pi q\int\limits_0^\infty du\; R^2(q,u)
\end{equation}
as an integral identity.

\section*{Appendix D: The function $J(q)$}
\setcounter{equation}{0}
\renewcommand{\theequation}{D.\arabic{equation}}

\noindent
Using again the method of Macke yields
\begin{eqnarray}\label{D1}
J(q\leq 2)=\frac{\pi}{4}q\left[\frac{8}{3}-4\ln 2 \right. 
&+&\frac{1}{3}\left(2-\frac{q}{2}\right)\left(1+\frac{2}{q}\right)^2\ln\left(1+ \frac{q}{2}\right) \nonumber \\
&+&\left.\frac{1}{3}\left(2+\frac{q}{2}\right)\left(1-\frac{2}{q}\right)^2\ln\left(1-\frac{q}{2}\right)\right], \nonumber \\
J(q\geq 2)=\frac{4\pi}{3}q\left[1\right.&+&\frac{1}{8q}(1-q)(2+q)^2\ln\left(1+\frac{2}{q}\right) \nonumber \\
&-&\left.\frac{1}{8q}(1+q)(2-q)^2\ln\left(1-\frac{2}{q}\right)\right].
\end{eqnarray}
$J(q)$ has the properties
\begin{eqnarray}\label{D2}
J(q\to 0)&=&\pi (1-\ln 2)\ q-\frac{\pi}{48}\ q^3+\cdots, \quad 
J(q\to \infty)=\frac{4 \pi}{3}\left(\frac{1}{q^2}+\frac{8}{15}\frac{1}{q^4}+\cdots\right), \nonumber \\
J(2)&=&\frac{4\pi}{3}(1-\ln 2), \quad J'(2^+)=-\frac{\pi}{3}(-1+4\ln 2), \quad J'(2^-)=-\frac{\pi}{6}(-5+8\ln 2). \nonumber \\  
\end{eqnarray}
(Notice $I(q\to 0)=\D\frac{8\pi}{3}J(q\to 0)$.)
$J(q)$ has a maximum of $1.3$ at $q=1.9$. $J(q)$ is continuous at $q=2$, but $J'(q)$ has there a 
jump discontinuity. Its normalization is 
\begin{eqnarray}\label{D3}
\int_0^\infty dq\; J(q)&=&\frac{\pi}{9}(-3+\pi^2+6\ln2). 
\end{eqnarray}
$J(q)$ is shown in Fig. 10. Multiplying the integral
\begin{eqnarray}\label{D4}
\int\limits_0^\infty\frac{dq}{q^2}\ [J(q)-\pi(1-\ln 2)\ q \Theta(1-q)]&=& 
\frac{\pi}{8}[10-\pi^2+8(1-\ln 2)\ln 2] \approx 0.719405
\end{eqnarray}
with $-2/\pi^3$ yields (\ref{c13}). \\

\noindent
The original expression for $J(q)$ arises from the diagram rules for $\Sigma_{2{\rm d}}$ with 
Eq. (\ref{c3}) as
\begin{equation}\label{D5}
J(q)={\rm Re}\ \int d^2e\int\frac{d\eta}{2\pi{\rm i}} Q(q,\eta)
\left[\frac{\Theta(|{\mbox{\bm $e$}}+{\mbox{\bm $q$}}|-1)}{{\mbox{\bm $q$}}({\mbox{\bm $e$}}+\frac{1}{2}{\mbox{\bm $q$}})-\eta -{\rm i}\delta}+\frac{\Theta(1-|{\mbox{\bm $e$}}+\mbox{\bm $q$}|)}{{\mbox{\bm $q$}}({\mbox{\bm $e$}}+\frac{1}{
2}{\mbox{\bm $q$}})-\eta +{\rm i}\delta}\right]
\end{equation} 
One way is to insert (\ref{A4}). It results 
[with $x_1={\mbox{\bm $q$}}({\mbox{\bm $e$}}_1+\frac{1}{2}{\mbox{\bm $q$}})$,
$x_2={\mbox{\bm $q$}}({\mbox{\bm $k$}}_2+\frac{1}{2}{\mbox{\bm $q$}})$]
\begin{equation}\label{D6}
J(q)=\int \frac{d^2e_1}{4\pi}d^3k_2
\left[\frac{\Theta(|{\mbox{\bm $e$}}_1+{\mbox{\bm $q$}}|-1)\ P}{x_1+x_2}
+\frac{\Theta(1-|{\mbox{\bm $e$}}_1+{\mbox{\bm $q$}}|)\ P}{x_1-x_2}\right]\Theta(1-k_2)
\Theta(|{\mbox{\bm $k$}}_2+{\mbox{\bm $q$}}|-1)
\end{equation}
in agreement with Eq. (\ref{c4}). Another way is the 
deformation of the integration contour from the real to the imaginary axis 
with $\eta={\rm i}qu$. It yields
\begin{equation}\label{D7} 
J(q)=
\int\limits_0^\infty du\; \left[\ln\frac{u^2+(1+\frac{q}{2})^2}{u^2+(1-\frac{q}{2})^2}\right]R(q,u)
\end{equation}
as an integral identity. \\

\noindent
Comparing $J(q)$ with $I(q)$:
\begin{equation}\label{D8}
\int\limits_0^\infty\frac{dq}{q^2}\left[\frac{3}{8\pi}I(q)-J(q)\right]=\frac{\pi^3}{3}a\ . 
\end{equation}
Note $\frac{3}{8\pi}I(q\to 0)=J(q\to 0)=\pi^3aq$. The identity (\ref{D8}) leads to the sum rule (\ref{c15}).

\end{appendix}
\begin{center}
{\bf Figures}
\end{center}

\begin{figure}
\includegraphics{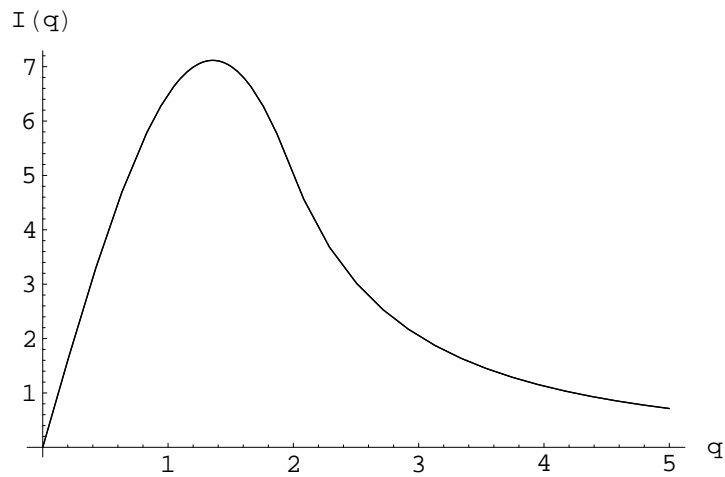}
\caption{The Macke function $I(q)$ according to Eq. (\ref{C1}), $I(q\to 0)=\frac{8\pi^4}{3}aq$.}
\end{figure}

\begin{figure}
\includegraphics{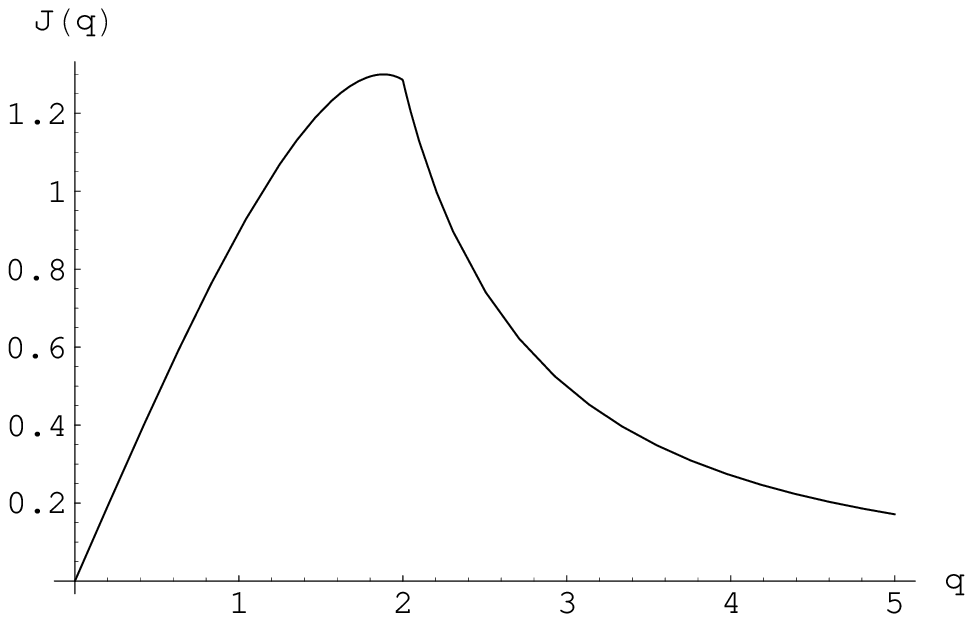}
\caption{The function $J(q)$ according to Eq. (\ref{D1}), $J(q\to 0)=\pi^3aq$.}
\end{figure}


\begin{thebibliography}{1}
\bibitem{Tos} M.P. Tosi in: N.H. March (Ed.), 
{\it Electron Correlation in the Solid State}, Imperial College Press, 1999, p. 1.
\bibitem{Zie1} There is a non-vanishing probability of finding 0, 2, 3, ... electrons in such a Wigner sphere.
Such particle number fluctuations in fragments have been studied by P. Ziesche, J. Tao, M. Seidl, and J.P. Perdew, 
Int. J. Quantum Chem. {\bf 77}, 819 (2000) with the conclusion `correlation suppresses fluctuations', 
cf. also P. Fulde, {\it Electron Correlations in Molecules and Solids}, 3rd ed., Springer, Berlin, 1995, p.157. 
\bibitem{foo} The Coulomb repulsion takes the form $\alpha r_s/r$, if lengths and energies are measured in units of
$k_{\rm F}^{-1}$ and $k_{\rm F}^2$, respectively. This shows that $r_s$ is not only a density parameter \cite{Zie1},
but also the interaction strength. Its Fourier transform is $4\pi\alpha r_s/q^2$ with $q$ measured in units of 
$k_{\rm F}$.
\bibitem{Zie2} P. Ziesche, Int J. Quantum Chem. {\bf 90}, 342 (2002).
\bibitem{Zie3} P. Ziesche and J. Cioslowski, Physica A {\bf 356}, 598 (2005). Here it is shown, which peculiarities 
of $n(k)$ and $S(q)$ caused by RPA lead to $e_{\rm c}\sim r_s^2\ln r_s$.
\bibitem{Cio} J. Cioslowski and P. Ziesche, submitted.
\bibitem{Mui} R.D. Mui\~no, I. Nagy, and P.M. Echenique, Phys. Rev. B {\bf 72}, 075117 (2005).
\bibitem{Hei} W. Heisenberg, Z. Naturf. {\bf 2a}, 185 (1947).
\bibitem{Ma}  W. Macke, Z. Naturf. {\bf 5a}, 192 (1950).
\bibitem{GB}  M. Gell-Mann and K. Brueckner, Phys. Rev. {\bf 106}, 364 (1957).
\bibitem{Ons} L. Onsager, L. Mittag, and M. J. Stephen, Ann. Physik (Leipzig) {\bf 18}, 71 (1966). Here the exchange 
integral of the 3D electron gas has been evaluated {\it analytically}. This rather herculean work was generalized and 
extended to the $d$-dimensional electron gas by M.L. Glasser \cite{Gla} 
\bibitem{Gla} M.L. Glasser, J. Comp. Appl. Math. {\bf 10}, 293 (1984).
\bibitem{GGZ} For a recent parametrization of $n(k,r_s)$ see P. Gori-Giorgi and P. Ziesche, 
Phys. Rev. B {\bf 66}, 235116 (2002). Its intention was to provide an {\it analytical} expression for $n(k,r_s)$ in
the range $r_s=1,\cdots,10$. In this range $n(k,r_s)$ is carefully fitted to the well-known kinetic energy $t(r_s)$. But numerical 
inaccurancies in the fitted expressions (17), (18), (19) made, that 
$n(k,r_s)$ is for $r_s=0,\cdots,1$ not accurate enough fo fully recover the well-known RPA behavior for 
${r_s{_> \atop ^{\to}}0}$.
\bibitem{Gal} W.M. Galitskii and A.B. Migdal, Zh. $\acute {\rm E}$ksp. Teor. Fiz. {\bf 34}, 139 (1958) 
[Sov. Phys. JETP {\bf 7}, 96 (1958)].
\bibitem{Mar} N.H. March, Phys. Rev. {\bf 110}, 604 (1958).
\bibitem{Lu} J.M. Luttinger, Phys. Rev. {\bf 121}, 942, 1251 (1961)
\bibitem{Hug} N.M. Hugenholtz and L. van Hove, Physica {\bf 24}, 363 (1958). Their theorem has been brought into 
the form $\mu-\mu_0=\Sigma(1,\mu)$ in ref. \cite{LuWa}. 
\bibitem{LuWa} J.M. Luttinger and J.C. Ward, Phys. Rev. {\bf 118}, 1417 (1960).
\bibitem{Sei} F. Seitz, {\it Modern Theory of Solids}, McGraw-Hill, New York, 1940, Sec. 76. 
\bibitem{Zie4} P. Ziesche, cond-mat/0605188 and Ann. Phys. (Leipzig), in press.  
\bibitem{foo2} The renormalized ring-diagram partial summation $\Sigma_{\rm r}^{\rm x}=
G_{\rm x}\cdot(r_{\rm r}-v_0)$ yields the correct $r_s\to 0$ asymptotics up to $r_s^2$, see Eqs. (\ref{d1}) and 
(\ref{d2}). Thus all the other remaining diagrams
or summations (e.g. $\Sigma_{\rm c}^{\rm HF}=(G-G_0)v_0$ of ref. \cite{Far}) can contribute only to 
higher-order asymptotics.  
\bibitem{Far} B. Farid, Phil. Mag.{\bf 84}, 109 (2004), Eq. (84). 
\bibitem{Ho} G.G. Hoffman, Phys.Rev. B{\bf 45}, 8730 (1992).
\bibitem{Da} E. Daniel and S.H. Vosko, Phys. Rev. {\bf 120}, 2041 (1960); 
I.O. Kulik, Zh. $\acute {\rm E}$ksp. teor. {\bf 40}, 1343 (1961) [Sov. Phys. JETP {\bf 13}, 946 (1961)]. 
\end{thebibliography}
\end{document}